\documentclass[12pt]{article}

\usepackage{amssymb}

\usepackage{graphics}
\usepackage{epsfig}

\newcommand{\nn}{\nonumber}

\newcommand{\bq}{\begin{eqnarray} }
\newcommand{\eq}{\end{eqnarray} }

\begin{document}

\begin{titlepage}

\begin{flushright}
OSU-HEP-03-14\\
December 2003\\
\end{flushright}
\vspace*{0.5cm}
\begin{center}
{\Large {\bf TeV--Scale Horizontal Symmetry and the \\[0.03in]
Slepton Mass Problem of Anomaly Mediation } }

\vspace*{1cm}
 {\large {\bf O.C. Anoka,$^{a,}$\footnote{E-mail address:anoka@okstate.edu}
 K.S. Babu$^{a,}$\footnote{E-mail address:  babu@okstate.edu}
 and I. Gogoladze$^{a,b,}$\footnote{On a leave of absence from:
Andronikashvili Institute of Physics, GAS, 380077 Tbilisi,
Georgia.  \\ E-mail address: gogoladze.1@nd.edu}}}

 \vspace*{1cm}
{\it $^a$ Department of Physics, Oklahoma State University\\
Stillwater, OK~74078, USA \\

$^b$ Department of Physics, University of Notre Dame \\Notre Dame,
IN 46556, USA}
\end{center}

 \vspace*{1.0cm}

\begin{abstract}

 We propose a new scenario for solving the tachyonic slepton mass problem of anomaly mediated supersymmetry
  breaking models with a non--Abelian horizontal gauge symmetry broken at
  the TeV scale. A specific model based on $SU(3)_{H}$ horizontal symmetry is presented
  wherein the sleptons receive positive mass--squared from the
  asymptotically free $SU(3)_{H}$ gauge sector. Approximate global symmetries present in
  the model strongly suppress flavor changing processes induced by the horizontal vector gauge bosons.
   The model predicts $m_{h}\lesssim 120$ GeV for the lightest
Higgs boson mass, $\tan{\beta}\simeq
 4$, and $M_V = 1-4$ TeV for the $SU(3)_{H}$ gauge boson masses.  The lightest SUSY particle is
 found to be the neutral Wino, which is a candidate for cold dark
 matter.

\end{abstract}

\end{titlepage}

\newpage

\section{Introduction}
Supersymmetry provides an elegant solution to the gauge hierarchy
problem of the standard model. To be realistic, it must however be
a broken symmetry. There are several ways of achieving
supersymmetry (SUSY) breaking. Anomaly mediated
  SUSY breaking (AMSB) is an attractive and predictive
 scenario which has the virtue that it can solve the SUSY flavor problem \cite{Randall,Giudice}.
This scenario assumes that SUSY breaking takes place in a hidden
or $sequestered$ sector. The MSSM superfields are confined to a
3--brane in a higher dimensional bulk space--time  separated from
the sequestered sector by extra dimensions. A rescaling
super--Weyl anomaly generates coupling of the auxiliary field of
the gravity multiplet to the gauginos and the scalars of the MSSM,
with the couplings determined by the SUSY renormalization group
equations (RGE). Since the rescaling anomaly is UV insensitive,
the pattern of SUSY breaking masses at any energy scale is
governed only by the physics at that scale [1--3]. Arbitrary
flavor structure in the SUSY scalar spectrum at high energies gets
washed out at low energies. This ultraviolet insensitivity
provides an elegant solution to the SUSY flavor problem.

 In anomaly mediated supersymmetry breaking
models, the masses of the scalar components of the chiral
supermultiplets are given by \cite{Randall,Giudice}
\begin{eqnarray}\label{AC1}
(m^2)^{\phi_{j}}_{\phi_{i}}=\frac{1}{2}M_{aux}^2\left[\beta(Y)\frac{\partial}{\partial{Y}}\gamma^{\phi_{j}}_{\phi_{i}}
+\beta(g)\frac{\partial}{\partial{g}}\gamma^{\phi_{j}}_{\phi_{i}}\right].
\end{eqnarray}
In the above equation  summations over the gauge couplings $g$ and
the Yukawa couplings $Y$ are assumed.
$\gamma^{\phi_{j}}_{\phi_{i}}$ are the one--loop anomalous
dimensions, $\beta (Y)$ is the beta function for the Yukawa
coupling $Y$, and $\beta (g)$ is the beta function for the gauge
coupling $g$. $M_{aux}$ is the vacuum expectation value of a
``compensator superfield" \cite{Randall} which sets the scale of
SUSY breaking. The gaugino mass $M_{g}$ associated with the gauge
group with coupling $g$ is given by \cite{Randall,Giudice}
\begin{eqnarray}\label{AC3}
M_{g}=\frac{\beta{(g)}}{g}M_{aux} .
\end{eqnarray}
The trilinear soft supersymmetry breaking term $A_{Y}$
corresponding to the Yukawa
 coupling $Y$ is given by \cite{Randall,Giudice}
\begin{eqnarray}\label{AC2}
A_{Y}=-\frac{\beta{(Y)}}{Y}M_{aux}.
\end{eqnarray}
 In the simplest scenario for generating the $\mu$ term, which we adopt in this paper, the
contribution to the  Higgs mixing parameter (the $B$-term) is
given by \cite{Randall}
\begin{eqnarray}\label{AN50}
B&=&-\left(\gamma_{H_{u}}+\gamma_{H_{d}}\right)M_{aux}.
\end{eqnarray}
The one--loop anomalous dimensions $\gamma_{H_{u}}$ and
$\gamma_{H_{d}}$ of  the $H_{u}$ and $H_{d}$ fields are given in
the Appendix (see Eqs. (45), (46)). Similar relations hold for
other bilinear terms in the SUSY breaking Lagrangian.

An attractive feature of the AMSB scenario is that it can
naturally solve the cosmological gravitino abundance  problem
which tends to destroy the success of big bang cosmology in
generic supergravity models \cite{Yamaguchi}. From Eqs.
(\ref{AC1}) and (\ref{AC3}) it follows that the masses of the
squarks, sleptons and the gauginos are all of order
$\frac{M_{aux}}{16\pi^2}$. The gravitino mass, on the other hand,
is $M_{gravitino}\sim M_{aux}$, which is naturally in the range 10
TeV -- 100 TeV in AMSB models (for MSSM sparticle masses of order
100 GeV -- 1 TeV). This is in contrast with $M_{gravitino}\sim
M_{SUSY}\sim$ TeV in generic supergravity models. With a mass in
the range 10 TeV -- 100 TeV, the gravitino lifetime is less than a
second, which suggests that the success of the big bang
nucleosynthesis will be preserved in AMSB scenario \cite{Fengm}.
Furthermore, the decay of the moduli fields present in the model
(as well as the gravitino) will produce neutralinos, especially
the neutral Winos, with the right abundance to make it a viable
cold dark matter candidate \cite{cold}.

In the minimal scenario, it turns out that AMSB induces negative
mass--squared for the sleptons. Such a scenario is excluded since
it would
 break electromagnetism. The reason for the negative mass--squared
 can be understood as follows.
There are two sources for slepton masses in AMSB, the Yukawa part
and the gauge part (Cf: Eq.. (\ref{AC1})). For the first two
families the Yukawa couplings are negligible and the dominant
contributions arise proportional to the gauge beta function. Since
in the minimal supersymmetric standard Model (MSSM) the
$SU(2)_{L}$ and the $U(1)_{Y}$ gauge couplings are not
asymptotically free, their gauge beta functions are positive. This
makes the slepton mass--squared negative. In the squark sector,
the masses are positive because $SU(3)_{C}$ gauge theory is
asymptotically free.

 Several possible ways of avoiding the slepton mass problem of AMSB have been
 suggested. A non--decoupling universal bulk contribution to all the scalar
 masses is a widely studied option [1--3, 5--8]. While this will make
 the minimal model phenomenologically consistent, the
 UV insensitivity of AMSB is no longer guaranteed. It is  therefore interesting to
 investigate variations of the minimal model which maintain the UV
 insensitivity but provide positive mass--squared for the sleptons
 from physics at the TeV scale. This is what we pursue in this
 paper.

One way to avoid the negative slepton mass problem with TeV scale
physics is to increase the Yukawa contributions in Eq.
(\ref{AC1}). This can be achieved by introducing new  particles at
the TeV scale with large Yukawa couplings to the lepton fields.
This possibility was studied in Ref. \cite{Chacko} where the MSSM
spectrum was extended to include 3 pairs of Higgs doublets, four
singlets and a vector--like pair of color--triplets near the weak
scale. The Yukawa contributions can also be enhanced by invoking
$R$--parity violating couplings in the MSSM \cite{Allanach1}.
Unfortunately such a theory would generate unacceptably large
neutrino masses. Yet another possibility is to make use of the
positive $D$--term contributions from a $U(1)$ gauge symmetry
broken at the weak scale. This was achieved by adding TeV scale
Fayet--Iliopoulos terms explicitly to the theory in Ref.
\cite{Jack}. New $D$--term contributions generated in a controlled
fashion by the breaking of $U(1)_{B-L}$ at an arbitrary high scale
may also provide positive contributions to the slepton masses
\cite{Arkani-Hamed,Hanik}.  A low scale ancillary $U(1)$ as a
solution to the problem has been studied in Ref. \cite{murakami}.
Effective supersymmetric theories which are devoid of the negative
slepton mass problem of AMSB with new dynamics at the 10 TeV scale
have been studied in Ref. \cite{Nelson}. Non--decoupling effects
of heavy fields at higher orders have been analyzed in AMSB models
in Ref. \cite{Katz} as an attempt to solve the slepton mass
problem.

The purpose of this paper is to suggest and investigate the
possibility of solving the negative slepton mass problem by making
the gauge contribution in  Eq. (\ref{AC1}) positive. This can only
be achieved by introducing a new non--Abelian gauge symmetry for
leptons with negative gauge beta function. We point out that an
$SU(3)_H$ horizontal symmetry  acting on the lepton multiplets has
all the desired properties for achieving this. We show that such
an $SU(3)_{H}$ horizontal symmetry broken at the TeV scale is
consistent with rare leptonic processes owing to the emergence of
approximate global symmetries.

The specific AMSB model we study is quite predictive. The lightest
Higgs boson mass is predicted to be $m_{h}\lesssim 120$ GeV, and
the parameter $\tan{\beta}$ is found to be $\tan{\beta}\simeq4$.
The model predicts the existence of new particles associated with
the $SU(3)_{H}$ symmetry breaking sector. The $SU(3)_{H}$ vector
bosons have masses of order 1--4 TeV. These particles should be
accessible experimentally at the LHC.

The plan of the paper is as follows. In Section 2 we introduce our
model. In section 3 we analyze the Higgs potential of the model.
Here we derive the limits on $\tan{\beta}$ and $m_h$. In section 4
we present the SUSY spectrum of the model and show how the
sleptons acquire positive masses. Numerical results for the full
spectrum of the model are given in Section 5. In Section 6 we
outline the most significant experimental consequences of the
model. In Section 7 we comment on the possible origins of the
$\mu$ and the $B\mu$ terms.   Section 8 has our conclusions. In an
Appendix we give the relevant beta functions, anomalous dimensions
as well as the soft masses.

\section{$SU(3)_{H}$ Horizontal Symmetry}

In this section we present our model. Since our aim is to have
positive contributions to the slepton masses from the gauge
sector, we are naturally led to a leptonic horizontal symmetry
that is asymptotically free. Our model is based on the gauge group
$SU(3)_{C}\otimes SU(2)_{L}\otimes U(1)_{Y}\otimes SU(3)_{H}$,
where $SU(3)_{H}$ is a horizontal symmetry acting on the leptons.
The left--handed lepton doublets and the antilepton singlets
transform as fundamental representations of the $SU(3)_{H}$ gauge
symmetry. The theory is made anomaly free by introducing three
Higgs multiplets ($\Phi_{1}$, $\Phi_{2}$, $\Phi_{3}$) which
transform as antifundamental representations of $SU(3)_H$ and as
singlets of the standard model. These fields are sufficient for
breaking the $SU(3)_H$ symmetry completely near the TeV scale. The
particle content of the model and the transformation properties
under the gauge group $SU(3)_{C}\otimes SU(2)_{L}\otimes
U(1)_{Y}\otimes SU(3)_{H}$ are presented  in Table 1. It turns out
that the Higgs potential involving these $\Phi_{i}$ fields
exhibits a global $SU(3)_{G}$ symmetry. We take advantage of this
global symmetry to suppress potentially large flavor changing
neutral current processes mediated by the $SU(3)_{H}$ gauge
bosons. The last column in Table 1 lists the transformation
properties under the global $SU(3)_{G}$ symmetry (The Yukawa
couplings of the model reduce the global $SU(3)_{G}$ down to
$U(1)$.) The fields $\eta_{i}$ and $\bar{\eta}_{i}$ are introduced
to facilitate $SU(3)_{H}$ symmetry breaking within our AMSB
framework.
\begin{table}[h]
\begin{center}
\begin{tabular}{|c|c|c|c|c|c|}\hline
\rule[5mm]{0mm}{0pt} Superfield &
 $ SU(3)_{C} $ & $SU(2)_{L}$& $U(1)_{Y}$ & $SU(3)_{H}$ & $SU(3)_{G}$\\
\hline \rule[5mm]{0mm}{0pt}$Q_{i}$&$3$&$2$&$\,\,\,\,\frac{1}{6}$&$1$&$1$\\
\hline
\rule[5mm]{0mm}{0pt}$u^c_i$&$\bar 3$&$1$&$-\frac{2}{3}$&$1$&$1$\\
\hline \rule[5mm]{0mm}{0pt}$d^c_i$&$\bar3$&1&$\,\,\,\,\frac{1}{3}$&$1$&$1$\\
\hline \rule[5mm]{0mm}{0pt}$L_{\alpha}$&$1$&$2$&$-\frac{1}{2}$&$3$&$1$\\
\hline \rule[5mm]{0mm}{0pt}$e^{c}_{\alpha}$&$1$&$1$&\,\,\,\,$1$&$3$&$1$\\
\hline \rule[5mm]{0mm}{0pt}$H_{u}$&$1$&$2$&$\,\,\,\,\frac{1}{2}$&$1$&$1$\\
\hline \rule[5mm]{0mm}{0pt}$H_{d}$&1&$2$&$-\frac{1}{2}$&$1$&$1$\\
\hline
\hline\rule[5mm]{0mm}{0pt}$\Phi^{\alpha}_{i}$&$1$&$1$&$0$&$\bar{3}$&$3$\\
\hline
\rule[5mm]{0mm}{0pt}$\eta_{i}$&$1$&$1$&$0$&$\bar{3}$&$3$\\
\hline \rule[5mm]{0mm}{0pt}$\bar{\eta}_{i}$&$1$&$1$&$0$&$3$&$\bar{3}$\\
\hline
\end{tabular}
\caption{\footnotesize Particle content and charge assignment of
the model. $SU(3)_{G}$ in the last column is a softly broken
global symmetry present in the model. The indices $i$ and $\alpha$
take values $i , \alpha$ = $1-3$.}
  \label{D0}
\end{center}
\end{table}

 Note that the
quarks are neutral under $SU(3)_{H}$. This is necessitated by the
requirements that $SU(3)_{H}$ be asymptotically free. A separate
$SU(3)_{H^{\prime}}$ acting on the quarks is a possible
quark--lepton symmetric extension of the model. But we do not
pursue such an extension here.

 The superpotential
of the model consistent with the gauge symmetries and the global
$SU(3)_G$ symmetry is given by:
\begin{eqnarray}\label{AN1}
W&=&\left(Y_{u}\right)_{ij}Q_{i}H_{u}u_{j}^{c}
+\left(Y_{d}\right)_{ij}Q_{i}H_{d}d_{j}^{c}+\mu H_{u}H_{d}\nn\\&+&
\kappa\Phi^{\alpha}_{1}\Phi^{\beta}_{2}\Phi^{\gamma}_{3}\epsilon_{\alpha\beta\gamma}
+\lambda
\eta_{a}^{\alpha}\eta_{b}^{\beta}\Phi_{c}^{\gamma}\epsilon_{\alpha\beta\gamma}\epsilon^{abc}
+M_{\eta}\eta_{a}{\bar{\eta}}_{a}.
\end{eqnarray}
Here $\alpha$, $\beta$, $\gamma$ =1, 2, 3  are $SU(3)_{H}$
indices, $i,j$ = 1, 2, 3 are family indices, and $a,b,c$ = 1, 2, 3
are $SU(3)_{G}$ indices. The mass parameters $\mu$ and $M_{\eta}$
are of order TeV, which has a natural origin in AMSB
\cite{Randall}. We will comment on possible origin of these terms
in Sec. 7.

 In the $SU(3)_{H}$ symmetric limit the leptons are all massless. They obtain their masses from the effective
 operators
\begin{eqnarray}\label{nn79}
L^{l}_{eff}=\frac{L_{\alpha}e^{c}_{\alpha}\Phi_{i}^{\alpha}\Phi_{i}^{\alpha}H_{d}}{M_{i}^2}.
\end{eqnarray}
Such operators can be obtained by integrating fields shown in Fig.
1, for example. The masses of the heavy fields break $SU(3)_{G}$
symmetry softly (the $\bar{\psi_i}\psi_i$ and the $\bar{E}_iE_i$
mass terms in Fig. 1).
\begin{figure}[!h]
\tiny{
\begin{center}
\includegraphics*[bb =-40 22 360 204 ]{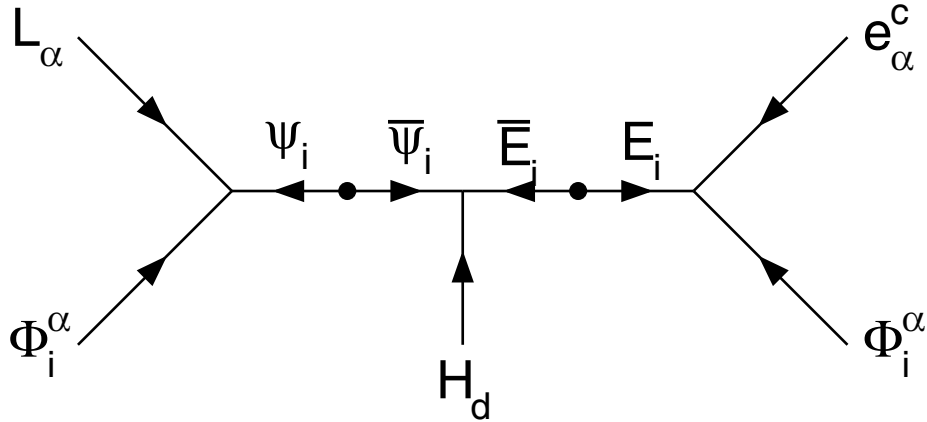}
\caption{ Effective operators inducing charged lepton masses.
}\label{fig1}
\end{center}}
\end{figure}Note that the mass scale $M_i$ in Eq. (\ref{nn79}) is of order 5
TeV for generating realistic $\tau$--lepton mass, of order 20 TeV
for the $\mu$ mass and of order 300 TeV for the electron mass
(assuming that all relevant Yukawa couplings are of order one).
Since these masses are all much heavier than the effective SUSY
breaking scale of order 1 TeV, these heavy fields will have no
effect in the low energy SUSY phenomenology within AMSB. Note that
no generation mixing is induced by these effective operators,
which will guarantee the approximate conservation of electron
number, muon number and tau lepton number. This is what makes the
model consistent with FCNC data even when $SU(3)_{H}$ is broken at
the TeV scale. Since the Higgs potential respects
$SU(3)_{H}\otimes SU(3)_{G}$ symmetry, after spontaneous symmetry
breaking, the diagonal subgroup $SU(3)_{G+H}$ remains as an
unbroken global symmetry. This subgroup contains $e$, $\mu$ and
$\tau$ lepton numbers.

Since right--handed neutrinos are not required to be light for
$SU(3)_{H}$ anomaly cancellation, they acquire heavy masses and
decouple from the low energy theory. Small neutrino masses are
then induced through the seesaw mechanism. Specifically, the
following effective nonrenormalizable operators emerge after
integrating out the heavy right--handed neutrino fields:
\begin{eqnarray}\label{nn7}
L^{\nu}_{eff}=\frac{\lambda_{ij}^{\alpha\beta}L_{\alpha}L_{\beta}H_uH_u
\Phi^{\alpha}_i\Phi^{\beta}_j}{M_N^3}.
\end{eqnarray}
Here $M_N$ represents the masses of the heavy right--handed
neutrino fields. For $M_N \sim 10^7$ GeV and
$\left\langle\Phi_{i}\right\rangle\sim$ TeV, neutrino masses in
the right range for oscillation phenomenology are obtained. Note
that Eq. (\ref{nn7}) arises from integrating neutral leptons with
their masses assumed to break all global symmetries. This enables
generation of large neutrino mixing angles, as needed for
phenomenology.

\section{Symmetry Breaking} The $SU(3)_{H}$ model has two sets of
Higgs bosons: the  usual MSSM Higgs doublets $H_{u}$ and $H_{d}$,
and the $SU(3)_{H}$ Higgs antitriplets $\Phi_{i}\,\,\ (i=1,2,3)$.
The Higgs potential is derived from the superpotential of Eq.
(\ref{AN1}) and includes the soft terms and the $D$ terms. The
tree level potential splits into two pieces:
\begin{eqnarray}\label{AN4}
V(H_{u},H_{d},\Phi_i)&=& V(H_{u},H_{d})+V(\Phi_i),
\end{eqnarray}
enabling us to analyze them independently. The first part,
$V(H_{u},H_{d})$, is identical to the MSSM potential which is well
studied. There are however significant constraints on the
parameters in our AMSB extension, which we now discuss.

\subsection{Constraints on $\tan{\beta}$ and $m_{h}$}
Minimization of $V(H_{u},H_{d})$ gives
\begin{eqnarray}\label{AN10}
\sin{2\beta}=
\frac{-2B\mu}{2\mu^2+m_{H_{u}}^2+m_{H_{d}}^2},\,\,\,\ \mu^2=
\frac{m_{H_{d}}^2-m_{H_{u}}^2\tan^2{\beta}}{\tan^2{\beta}-1}-\frac{M_{Z}^2}{2}.
\end{eqnarray}
Here $m_{H_{u}}^2$ and $m_{H_{d}}^2$ are the Higgs soft masses and
are given in the Appendix for the AMSB model (see Eqs.
(58)--(59).) The constraints on $m_h$ and $\tan{\beta}$ arise
since these soft masses and the $B$ parameter are determined in
terms of a single parameter $M_{aux}$ in our framework.

We eliminate $M_{aux}$ in favor of $M_{2}$, the Wino mass
($M_{2}=\frac{b_{2}g^2_{2}}{16\pi^2}M_{aux}$). We see from Eqs.
(\ref{AN10}), (\ref{AN50}) as well as from Eqs. (45)--(46) and
Eqs. (58)--(59) of the Appendix that $\tan{\beta}$ is fixed in
terms of $M_{2}$. In Fig. 2 we plot $\tan{\beta}$ as a function of
$M_{2}$. For the experimentally interesting range of
$M_{2}\gtrsim$ 100 GeV, we find that $\tan{\beta}\simeq$ 3.8 --
4.0.
 In obtaining the limit
on $\tan{\beta}$, we followed the following procedure. As inputs
at $M_{Z}$ we chose \cite{Particle}
\begin{eqnarray}\label{AN458}
\alpha_{3}(M_Z)=0.119,\,\,\,\,\sin^{2}{\theta_{W}}=0.2312,\,\,\,\,\alpha(M_Z)=\frac{1}{127.9}.
\end{eqnarray}
Using the central value of $M_{t}=174.3$ GeV, we obtain the
running mass $m_{t}(M_t)$ with the 2--loop QCD correction as
\cite{Ramond}
\begin{eqnarray}\label{AN451}
\frac{M_{t}}{m_{t}(M_t)}=1+\frac{4}{3}\frac{\alpha_3(M_t)}{\pi}+10.9\left(\frac{\alpha_3(M_t)}{\pi}\right)^2.
\end{eqnarray}

\begin{center}
\begin{figure}[!h]
\includegraphics*[bb = 59 230 503 557 ]{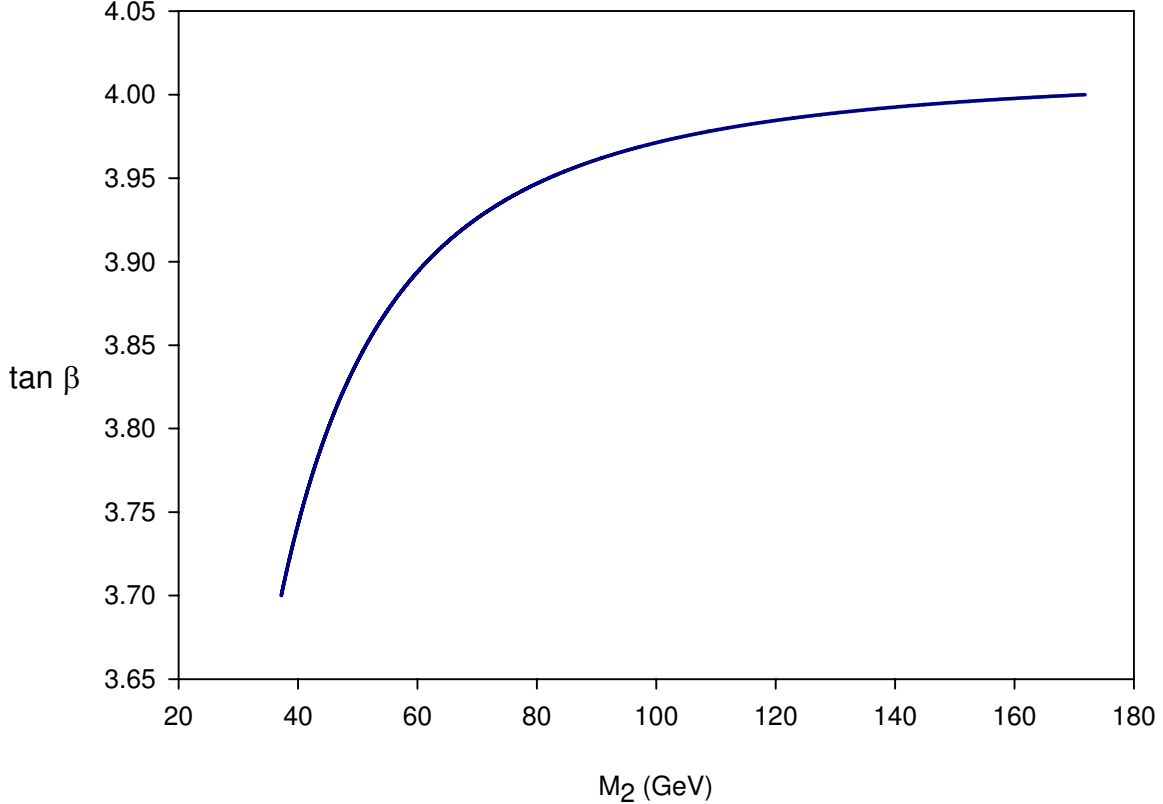}
\caption{ Plot of $\tan{\beta}$ as a function of
$M_{2}$}\label{fig2}
\end{figure}
\end{center}
Using 5--flavor SM QCD beta functions we extrapolated
$\alpha_{3}(M_Z)$ and obtained $\alpha_{3}(M_t)$ = 0.109. The top
quark Yukawa coupling is then found to be (for $M_{t}$ = 174.3
GeV) $Y^{SM}_t(M_t)$ = 0.935 corresponding to $m_t(M_t)=162.8$
GeV. This coupling is then evolved from $M_{t}$ to 1 TeV where we
minimize the MSSM Higgs potential. Using standard model beta
function we obtain $Y^{SM}_t(1 $ TeV$)$ = 0.851. The corresponding
MSSM coupling is $Y_t(1 $ TeV$) = Y^{SM}_t(1 $ TeV$)/\sin{\beta}$
, which for $\tan{\beta}\simeq 4.0$ (justified a--posteriori) is
$Y_t(1 $ TeV$) = 0.824$. The gauge couplings at 1 TeV are found to
be $g_{1}(1 $ TeV$) = 0.466,\,\,\,\,g_{2}(1$ TeV$) =
0.642\,\,\,\,$and$\,\,\,\,g_{3}(1 $ TeV$) = 1.098$. With these
values of couplings at 1 TeV we obtained Fig. 1. Uncertainties in
the prediction for $\tan{\beta}$ are estimated to be $\pm 0.5$,
arising from the error in top quark mass and from the precise
scale at which the Higgs potential is minimized. We conclude that
$\tan{\beta}$ = 3.5--4.5 in this model.

Since $\tan{\beta}$ is fixed and since the $A_{t}$ parameter is
not free in AMSB, there is a nontrivial prediction for the
lightest Higgs boson mass $m_{h}$. We use the 2--loop radiatively
corrected expression for $ m^2_{h}=(m^2_{h})_{o}+\Delta m_{h}^2$,
where $(m^2_{h})_{o}$ is the tree--level value of the mass and the
radiative correction is given by \cite{Carena2}
\begin{eqnarray}\label{AN133} \Delta
m_{h}^2&=&\frac{3m_{t}^{4}}{4\pi^2\upsilon^{2}}\left[t+X_{t}+\frac{1}{16\pi^2}\left(\frac{3}{2}\frac{m_{t}^2}{\upsilon^2}-32\pi\alpha_{3}(M_{t})\right)(2X_{t}t+t^2)\right].
 \end{eqnarray}
Here
 \begin{eqnarray}\label{AN134}
 X_{t}=
\frac{\tilde{A_{t}}^2}{m_{\tilde{t}}^{2}}\left(1-\frac{\tilde{A_{t}}^2}{12m_{\tilde{t}}^{2}}\right),\,\,\,\
\tilde{A_{t}}=A_{t}-\mu\cot{\beta},
\end{eqnarray}
and $ t=$log$(\frac{m_{\tilde{t}}^{2}}{M_{t}^{2}})$,
$\upsilon=174$ GeV. $m_{\tilde{t}}^{2}$ is the arithmetic average
of the diagonal entries of the squared stop mass matrix and
$A_{t}$ is the soft trilinear coupling associated with the top
Yukawa coupling in the superpotential of Eq. (\ref{AN1}). In these
expressions, $m_{t}$ is the one--loop QCD corrected running mass,
$m_t=\frac{M_t}{1+\frac{4}{3}\frac{\alpha_3(M_t)}{\pi}}$, which
equals 166.7 GeV for $M_{t} = 174.3$ GeV. We find that $m_h\simeq
$ 113 GeV -- 120 GeV, depending on the choice of $M_{aux}$. The
larger value $m_h\simeq 120$ GeV is realized only for larger
$M_{t}\simeq 180 $ GeV. We list in Tables 2--4 the value of $m_h$,
along with the other sparticle masses.

\subsection{$SU(3)_{H}$ symmetry breaking}

Let us now analyze the $SU(3)_H$ symmetry breaking sector of the
potential. The potential $V(\Phi_i)$ is given by:
\begin{eqnarray}\label{AN17}
V(\Phi_i)&=&
m_{\phi}^2(\Phi_{1}^{\dag}\Phi_{1}+\Phi_{2}^{\dag}\Phi_{2}+\Phi_{3}^{\dag}\Phi_{3})+\kappa
A_{\kappa}\left(\Phi_{1}^{\alpha}\Phi_{2}^{\beta}\Phi_{3}^{\gamma}\epsilon_{\alpha\beta\gamma}+c.c\right)\nn\\&+&\kappa^{2}
\left[(\Phi_{1}\Phi_{2})^{\dag}(\Phi_{1}\Phi_{2})+(\Phi_{1}\Phi_{3})^{\dag}(\Phi_{1}\Phi_{3})+(\Phi_{2}\Phi_{3})^{\dag}(\Phi_{2}\Phi_{3})\right]\nn\\&+&
\frac{g_{4}^{2}}{8}\sum_{a=1}^{8}|\Phi_{1}^{\dag}\lambda^{a}\Phi_{1}+
\Phi_{2}^{\dag}\lambda^{a}\Phi_{2}+\Phi_{3}^{\dag}\lambda^{a}\Phi_{3}|^{2}.
\end{eqnarray}
Here $g_{4}$ is the gauge coupling of the $SU(3)_{H}$,
$A_{\kappa}$ is the trilinear $A$--term corresponding to the
coupling $\kappa$,  $m^{2}_{\phi}$ is the soft mass squared for
the $\Phi_i$ fields. These soft SUSY breaking parameters are given
in the Appendix (Eqs. (56), (60)). The $\kappa^2$ term in Eq.
(\ref{AN17}) is the $F$-term contribution and the last term in Eq.
(\ref{AN17}) is the $SU(3)_{H}$ $D$--term with $\lambda^{a}$ being
the $SU(3)_H$ generators.

The Higgs potential, Eq. (\ref{AN17}), has an $SU(3)_{H}\times
SU(3)_{G}$ symmetry, with the $\Phi_i$ fields $(i=1-3)$
transforming as $\bf{(\bar{3},3)}$. This allows for a vacuum which
preserves an $SU(3)_{H+G}$ diagonal subgroup. The VEVs of the
$\Phi_i$ fields are then given by:

\begin{eqnarray}\label{nn71}
\left\langle\Phi_{1}\right\rangle=\pmatrix{u\cr 0\cr0},\qquad
\qquad \left\langle\Phi_{2}\right\rangle=\pmatrix{0\cr u
\cr0}\qquad \mbox{and} \qquad
\left\langle\Phi_{3}\right\rangle=\pmatrix{0\cr 0\cr u}.
\end{eqnarray}
Using these VEVs the potential becomes
\begin{eqnarray}\label{AN18}
\left\langle
V(\Phi)\right\rangle&=&3m_{\phi}^{2}u^{2}+3\kappa^{2}u^{4}+2\kappa
A_{\kappa}u^{3}.
\end{eqnarray}
 Minimization of Eq. (\ref{AN18}) leads to the condition
\begin{eqnarray}\label{AN20}
u&=&\frac{-
A_{\kappa}\pm\sqrt{-8m_{\phi}^{2}+A_{\kappa}^{2}}}{4\kappa}.
\end{eqnarray}
The argument in the square root of Eq. (\ref{AN20}), which should
be positive for a consistent symmetry breaking, is given by
\begin{eqnarray}\label{AN233}
-8m_{\phi}^{2}+A_{\kappa}^{2}& =
&\frac{M_{aux}^2}{(16\pi^2)^2}[15\kappa^4+56\kappa^2\lambda^2+304\lambda^4-8\kappa^2g_{4}^2-32\lambda^2g_{4}^2].
\end{eqnarray}
Positivity of Eq. (\ref{AN233}) leads to constraints on the
parameters
 $\{\lambda,\,\,\kappa\}$. It can be shown that Eq. (\ref{AN233}) implies
$0\leqslant|\kappa|\leqslant 0.731 g_{4}$ and
$0\leqslant|\lambda|\leqslant 0.324g_{4}$. Furthermore, positivity
of the slepton masses, along with the experimental limit
$m^2_{slepton} \gtrsim (100 $ GeV$)^2$, require $g_{4}\geqslant
0.5$. This essentially fixes the parameter space of the model. We
get the right minimum by choosing the negative sign of the square
root in Eq. (\ref{AN20}) (for positive $M_{aux}$), with this
choice, all the Higgs masses--squared will be positive.

Since the symmetry breaking chain is $SU(3)_{H}\times
SU(3)_G\rightarrow SU(3)_{H+G}$, we can classify the masses of all
scalars and fermions as multiplets of $SU(3)_{H+G}$. The complex
$\Phi(\bar{3},3)$ scalar multiplet decomposes into 2 octets and
two singlets of $SU(3)_{H+G}$. One octet gets eaten by the Higgs
mechanism. A physical octet remains in the spectrum with a mass
given by
\begin{eqnarray}\label{AN202}
\mathcal{M}^2_{octet}&=&-2\kappa^{2}u^{2}-2\kappa
uA_{\kappa}+g_{4}^{2}u^{2}.
\end{eqnarray}
There are two singlets, one scalar ($\phi_s$) and one pseudoscalar
($\phi_p$) with masses given by
\begin{eqnarray}\label{AN203}
m^2_{\phi_{s}}&=&4\kappa^{2}u^{2}+\kappa u A_{\kappa},\\
m^2_{\phi_{p}}&=&-3\kappa uA_{\kappa}.
\end{eqnarray}

In the fermionic sector, the octet Higgsino mixes with the octet
gaugino with a mixing matrix given by
\begin{eqnarray}\label{AN200}
\mathcal{M}_{octet}^{'}&=&\pmatrix{m_{4}&g_{4}u \cr g_{4}u& \kappa
u}.
\end{eqnarray}
In addition, there is a Majorana fermion, a singlet of
$SU(3)_{H+G}$, with a mass of
\begin{eqnarray}\label{AN201}
m_{\tilde{{\phi}}}&=&2\kappa u.
\end{eqnarray}

Finally the gauge bosons form an octet with a mass
\begin{eqnarray}\label{AN240}
M_{V}=g_{4}u.
\end{eqnarray}

\section{The SUSY Spectrum}
We are now ready to discuss the full SUSY spectrum of the model.
We will see that the tachyonic slepton problem is cured by virtue
of the positive contribution from the $SU(3)_{H}$ gauge sector.
\subsection{Slepton masses}
 The slepton mass--squareds are given by the
eigenvalues of the mass matrices ($\alpha=e,\,\mu,\,\tau$)
\begin{eqnarray}\label{MM99}
M^2_{\tilde{l}} =\pmatrix{m^2_{\tilde{L}_\alpha} &
m_{E_{\alpha}}\left(A_{Y_{E_{\alpha}}}-\mu\tan{\beta}\right)\cr
m_{E_{\alpha}}\left(A_{Y_{E_{\alpha}}}-\mu\tan{\beta}\right)&
m^2_{\tilde{e}^{c}_\alpha}}.
\end{eqnarray}
Here
\begin{eqnarray}\label{AN2}
m^2_{\tilde{L}_\alpha}&=&\frac{M_{aux}^2}{(16\pi^2)}\left[Y_{E_{\alpha}}\beta(Y_{E_{\alpha}})-\left(\frac{3}{2}g_{2}\beta(g_{2})
+\frac{3}{10}g_{1}\beta(g_{1})+\frac{8}{3}g_{4}\beta(g_{4})\right)\right]\nn\\&+&m_{E_{\alpha}}^2+\left(-\frac{1}{2}+\sin^2{\theta_{W}}\right)\cos{2\beta}M_{Z}^{2},\\
m^2_{\tilde{e}^{c}_\alpha}&=&\frac{M_{aux}^2}{(16\pi^2)}\left[2Y_{E_{\alpha}}\beta(Y_{E_{\alpha}})-\left(\frac{6}{5}g_{1}\beta(g_{1})+\frac{8}{3}g_{4}\beta(g_{4})\right)\right]\nn\\
&+&m_{E_{\alpha}}^2-\sin^2{\theta_{W}}\cos{2\beta}M_{Z}^{2}.
\end{eqnarray}
The off diagonal terms in Eq. (\ref{MM99}) are rather small as
they are proportional to the lepton masses. The SUSY soft masses
are calculated from the RGE give in the Appendix. The last terms
of Eqs. (26)--(27) are the $D$--terms. Note the positive
contribution from the $SU(3)_{H}$ gauge sector in Eqs. (26)--(27),
given by the term $-\frac{8}{3}g_{4}\beta(g_{4})$. In our model
$g_4$ is asymptotically free with
$\beta(g_{4})=-\frac{3}{16\pi^2}g_{4}^3$. This contribution makes
the mass--squared of all sleptons to be positive for $g_4
\geqslant 0.5$.

\subsection{Squark  masses}
The mixing matrix for the squark sector is similar to the slepton
sector. The diagonal entries of the up and the down squark mass
matrices are given by \cite{Martin}
\begin{eqnarray}\label{AN33}
m_{\tilde{U}_{i}}^2&=&(m^{2}_{soft})_{\tilde{Q}_{i}}^{\tilde{Q}_{i}}+m_{U_{i}}^{2}+\frac{1}{6}\left(4M_{W}^{2}-M_{Z}^{2}\right)\cos{2\beta},\nn\\
m_{\tilde{U}^{c}_{i}}^2&=&(m^{2}_{soft})_{\tilde{U}_{i}^{c}}^{\tilde{U}_{i}^{c}}+m_{U_{i}}^{2}-\frac{2}{3}\left(M_{W}^{2}-M_{Z}^{2}\right)\cos{2\beta},\nn\\
m_{\tilde{D}_{i}}^2&=&(m^{2}_{soft})_{\tilde{Q}_{i}}^{\tilde{Q}_{i}}+m_{D_{i}}^{2}-\frac{1}{6}\left(2M_{W}^{2}+M_{Z}^{2}\right)\cos{2\beta},\nn\\
m_{\tilde{D}_{i}^{c}}^2&=&(m^{2}_{soft})_{\tilde{D}_{i}^{c}}^{\tilde{D}_{i}^{c}}+
m_{D_{i}}^{2}+\frac{1}{3}\left(M_{W}^{2}-M_{Z}^{2}\right)\cos{2\beta}.
\end{eqnarray}
Here $m_{U_{i}}$ and $m_{D_{i}}$ are quark masses of different
generations, $i$ = 1, 2, 3. The squark soft masses are obtained
from the RGE as
\begin{eqnarray}\label{A.27}
(m^{2}_{soft})_{\tilde{Q}_{i}}^{\tilde{Q}_{i}}=\frac{M_{aux}^{2}}{16\pi^2}\left(Y_{u_{i}}
\beta{(Y_{u_{i}})}+Y_{d_{i}}\beta{(Y_{d_{i}})}-\frac{1}{30}g_{1}\beta{(g_{1})}-\frac{3}{2}g_{2}
\beta{(g_{2})} -\frac{8}{3}g_{3}\beta{(g_{3})}\right),
\end{eqnarray}
\begin{eqnarray}\label{A.28}
(m^{2}_{soft})_{\tilde{U}_{i}^{c}}^{\tilde{U}_{i}^{c}}&=&\frac{M_{aux}^{2}}{16\pi^2}\left(2Y_{u_{i}}
\beta{(Y_{u_{i}})}-\frac{8}{15}g_{1}\beta{(g_{1})}-\frac{8}{3}g_{3}\beta{(g_{3})}\right),\\
(m^{2}_{soft})_{\tilde{D}_{i}^{c}}^{\tilde{D}_{i}^{c}}&=&\frac{M_{aux}^{2}}{16\pi^2}\left(2Y_{d_{i}}
\beta{(Y_{d_{i}})}-\frac{2}{15}g_{1}\beta{(g_{1})}-\frac{8}{3}g_{3}\beta{(g_{3})}\right).
\end{eqnarray}

\subsection{$\eta$  fermion and $\eta$ scalar masses }
The fields $\eta$ and $\bar{\eta}$ transform as $(3,\bar{3})$ and
$(\bar{3},3)$ under $SU(3)_{H}\times SU(3)_G$. After symmetry
breaking, $\eta$ and $\bar{\eta}$ both transform as ${\bf 8+1}$ of
the diagonal $SU(3)_{H+G}$. The octet from $\eta$ mixes with the
octet from $\bar{\eta}$, and similarly for the singlets.

In the fermionic sector, the octet and the singlet mass matrices
are given by
\begin{eqnarray}\label{A.229}
&&M_{octet}^{\eta}=\pmatrix{-2\lambda u & M_{\eta}\cr M_{\eta} & 0},\\
&&M_{singlet}^{\eta}=\pmatrix{4\lambda u & M_{\eta}\cr M_{\eta} &
0}.
\end{eqnarray}

In the scalar sector, there are 4 real octets and 4 real singlets
from $\eta$ and $\bar{\eta}$ fields. The two scalar octets are
mixed, as are the two pseudoscalar octets. The mass squared
matrices for the octet are

\begin{eqnarray}\label{A.232}
M^2_{s-octet}&=&\pmatrix{(\tilde{m}^{2}_{soft})_{\eta}^{\eta}+
M_{\eta}^2+2\lambda u(-A_{\lambda}-\kappa u +2\lambda u)&
M_{\eta}(B_{\eta}-2\lambda u )\cr  M_{\eta}(B_{\eta}-2\lambda u )
&(\tilde{m}^{2}_{soft})_{\bar{\eta}}^{\bar{\eta}}+M_{\eta}^2},
\end{eqnarray}
\begin{eqnarray}\label{A.2312}
M^2_{p-octet}&=&\pmatrix{(\tilde{m}^{2}_{soft})_{\eta}^{\eta}+
M_{\eta}^2+2\lambda u(A_{\lambda}+\kappa u +2\lambda u)&
-M_{\eta}(B_{\eta}+2\lambda u )\cr  -M_{\eta}(B_{\eta}+2\lambda u
) &(\tilde{m}^{2}_{soft})_{\bar{\eta}}^{\bar{\eta}}+M_{\eta}^2}.
\end{eqnarray}
The singlet scalar mass matrices are
\begin{eqnarray}\label{A.231}
M^2_{s-singlet}&=&\pmatrix{(\tilde{m}^{2}_{soft})_{\eta}^{\eta}+
M_{\eta}^2+4\lambda u(A_{\lambda}+\kappa u +4\lambda u)&
M_{\eta}(B_{\eta}+4\lambda u )\cr  M_{\eta}(B_{\eta}+4\lambda u )
&(\tilde{m}^{2}_{soft})_{\bar{\eta}}^{\bar{\eta}}+M_{\eta}^2},
\end{eqnarray}
\begin{eqnarray}\label{A.2311}
M^2_{p-singlet}&=&\pmatrix{(\tilde{m}^{2}_{soft})_{\eta}^{\eta}+
M_{\eta}^2-4\lambda u(A_{\lambda}-\kappa u -4\lambda u)&
-M_{\eta}(B_{\eta}-4\lambda u )\cr -M_{\eta}(B_{\eta}-4\lambda u )
&(\tilde{m}^{2}_{soft})_{\bar{\eta}}^{\bar{\eta}}+M_{\eta}^2}.
\end{eqnarray}
The soft masses $(\tilde{m}^{2}_{soft})_{\eta}^{\eta}$ and
$(\tilde{m}^{2}_{soft})_{\eta}^{\eta}$ are given in Eqs.
(61)--(62) of the Appendix.

\section{Numerical Results}
We are now ready to present our numerical results for the SUSY
spectrum. The scale of SUSY breaking, $M_{aux}$, should be in the
range 40--120 TeV for the MSSM particles to have masses in the
range 100 GeV -- 2 TeV. Note that there is a large hierarchy in
the masses of the gluino and the neutral Wino,
$\frac{M_3}{M_2}\simeq 7.1$ (after taking account of radiative
corrections), in AMSB models. Furthermore the lightest chargino is
nearly mass degenerate with the neutral Wino, so $M_2\gtrsim 100$
GeV is required to satisfy the LEP chargino mass bound.

The $SU(3)_{H}$ gauge coupling $g_4$ is chosen so that the
sleptons have positive mass squared ($g_4\geqslant 0.5$). We allow
$g_4$ to take two values, $g_4$ = 0.55 (Tables 2 and 4) and $g_4$
= 1.0 (Table 3). Symmetry breaking considerations constrain the
couplings $\kappa$ and $\lambda$ as discussed in Sec. 3 after Eq.
(18). In Tables 2 and 4 we have taken $M_{aux}$ = 47.112 TeV
corresponding to a light spectrum, while in Table 3 we have
$M_{aux}$ = 66.695 TeV with an intermediate spectrum. Other input
parameters are listed in the respective Table captions. The mass
parameter $M_{\eta}$ cannot be much larger than 1 TeV, as that
would decouple the effects of $\eta$, $\bar{\eta}$ fields which
are needed for consistent symmetry breaking.

We see from Table 2 that the lightest Higgs boson mass is $m_h
\simeq 113$ GeV. This is very close to the current experimental
limit. If $M_t$ = 180 GeV is used (instead of $M_t$ = 176 GeV),
for the same set of input parameters, $m_h$ will be 115 GeV. $m_h$
being close to the current experimental limit is a generic
prediction of our framework. It holds in the spectra of Tables 3
and 4 as well. We conclude that $m_h\lesssim 120$ GeV in this
model.

The masses of the sleptons will depend sensitively on the choice
of $g_4$. The sleptons are relatively light, $m_{slep}\lesssim$
300 GeV, with $g_4=0.55$, while they are heavy, $m_{slep}\simeq$
800 GeV, when $g_4=1.0$. Note however that there is a correlation
in the slepton masses and the $SU(3)_H$ gauge boson masses
($M_V$), with the lighter sleptons corresponding to lighter
$SU(3)_H$ gauge bosons. It is worth noting that very light
sleptons, below the current experimental limits of about 100 GeV,
would be inconsistent with the limits on $M_V$ arising from
$e^+e^-\rightarrow \mu^+\mu^-$ type processes (see Sec. 6). Note
also that the left--handed and the right--handed sleptons are
nearly degenerate to within about 10 GeV in this model.  This a
numerical coincidence having to do with the values of $g_1$ and
$g_2$ and the MSSM beta functions (see the last paper of Ref.
[5]). The new $SU(3)_H$ gauge boson contributions to the slepton
masses are the same for the left--handed and the right--handed
sleptons.

In Tables 2--4 we have included the leading radiative corrections
to the gaugino masses $M_1$, $M_2$ and $M_3$ \cite{Gunion}.
Including these radiative corrections we find (in Table 2)
$M_1:M_2:M_3=2.9:1:7.3$. The lightest SUSY particle (LSP) is the
neutral Wino, which is nearly mass degenerate with the charged
Wino. In Tables 2--4 the mass splitting is about 60 MeV, but this
does not take into account $SU(2)_L\times U(1)_Y$ breaking
corrections \cite{Pierce}. These electroweak radiative corrections
turn out to be very important, and we find
$m_{\chi^{\pm}_{1}}-m_{\chi_{1}^{0}}\simeq 235$ MeV (with about
175 MeV arising from $SU(2)_L\times U(1)_Y$  breaking effects).
The decay $\chi_{1}^{\pm}\rightarrow \chi_{1}^0+\pi^{\pm}$ is then
kinematically allowed, with the $\pi^{\pm}$ being very soft. Once
produced, the neutralino $\chi_1^0$ will escape the detector
without leaving any tracks.  With the decay channel $\chi_1^\pm
\rightarrow \chi_1^0 + \pi^\pm$ open, the lightest chargino will
leave an observable track with a decay length of about a few cm.
Search strategies for such a quasi--degenerate pair at colliders
have been analyzed in Ref. \cite{Gunion,wells,roy}.

In the $SU(3)_H$ sector, in Tables 2--4, the horizontal gauge
boson has a mass of 1.5--4.0 TeV. The heavy Higgs bosons,
Higgsinos, gauginos, squarks  and the $\eta$ fields all have
masses $\lesssim (1-2)$ TeV.
\newpage
~
 \vspace{0.05in}
\begin{center}
\begin{table}[!h]
\begin{center}
\begin{tabular}{|l|c|c|}\hline
\rule[4.0mm]{0mm}{0pt} MSSM Particles & Symbol& Mass (TeV)\\\hline
\rule[4.0mm]{0mm}{0pt}Neutralinos&$\{m_{\tilde{\chi}_{1}^{0}},\,\,m_{\tilde{\chi}_{2}^{0}},\,\,m_{\tilde{\chi}_{3}^{0}},\,\,m_{\tilde{\chi}_{4}^{0}}\}$&$\{0.14158,\,\,0.429,\,\,0.872,\,\,0.879\}$\\\hline
\rule[4.0mm]{0mm}{0pt}Charginos&$\{m_{\tilde{\chi}_{1}^{\pm}},\,\,m_{\tilde{\chi}_{2}^{\pm}}\}$&$\{0.14164,\,\,0.878\}$\\\hline
\rule[4.0mm]{0mm}{0pt}Gluino&$M_{3}$&$1.065$\\\hline
\rule[4.0mm]{0mm}{0pt}Higgs bosons
&$\{m_{h},\,\,m_{H},\,\,m_{A},\,\,m_{H^{\pm}}\}$&$\{0.113,\,\,0.897,\,\,0.896,\,\,0.900\}$\\\hline
\rule[4.0mm]{0mm}{0pt}R.H sleptons
&$\{m_{\tilde{e}_{R}},\,\,m_{\tilde{\mu}_{R}},\,\,m_{\tilde{\tau}_{1}}\}$&$\{0.183,\,\,0.183,\,\,0.170\}$\\\hline
\rule[4.0mm]{0mm}{0pt}L.H sleptons
&$\{m_{\tilde{e}_{L}},\,\,m_{\tilde{\mu}_{L}},\,\,m_{\tilde{\tau}_{2}}\}$&$\{0.190,\,\,0.190,\,\,0.200\}$\\\hline
\rule[4.0mm]{0mm}{0pt}Sneutrinos&$\{m_{\tilde{\nu}_{e}},\,\,m_{\tilde{\nu}_{\mu}},\,\,m_{\tilde{\nu}_{\tau}}\}$&$\{0.175,\,\,0.175,\,\,0.175\}$\\\hline
\rule[4.0mm]{0mm}{0pt}R.H down squarks
&$\{m_{\tilde{d}_{R}},\,\,m_{\tilde{s}_{R}},\,\,m_{\tilde{b}_{1}}\}$&$\{1.017,\,\,1.017,\,\,1.014\}$\\\hline
\rule[4.0mm]{0mm}{0pt}L.H down squarks
&$\{m_{\tilde{d}_{L}},\,\,m_{\tilde{s}_{L}},\,\,m_{\tilde{b}_{2}}\}$&$\{1.008,\,\,1.008,\,\,0.886\}$\\\hline
\rule[4.0mm]{0mm}{0pt}R.H up squarks
&$\{m_{\tilde{u}_{R}},\,\,m_{\tilde{c}_{R}},\,\,m_{\tilde{t}_{1}}\}$&$\{1.011,\,\,1.011,\,\,0.718\}$\\\hline
\rule[4.0mm]{0mm}{0pt}L.H up squarks
&$\{m_{\tilde{u}_{L}},\,\,m_{\tilde{c}_{L}},\,\,m_{\tilde{t}_{2}}\}$&$\{1.005,\,\,1.005,\,\,0.944\}$\\
\hline\hline \rule[4.0mm]{0mm}{0pt} New Particles&Symbol& Mass
(TeV)\\\hline\hline
 \rule[4.0mm]{0mm}{0pt}$SU(3)_{H}$ Gauge boson
octet&$M_{V}$&$2.213$\\\hline
 \rule[4.0mm]{0mm}{0pt}Singlet Higgsino
&$m_{\tilde{{\phi}}}$&$0.402$\\\hline \rule[4.0mm]{0mm}{0pt}Octet
Higgsino/gaugino
&$m_{\tilde{\phi}_{{1,2}}}$&$\{1.978,\,\,2.450\}$\\\hline
\rule[4.0mm]{0mm}{0pt}$\phi$ Higgs bosons
&$\{m_{\phi_{s}},m_{\phi_{p}},m_{\phi-octet}\}$&$\{0.179,\,\,0.624,\,\,2.253\}$\\\hline
\rule[4.0mm]{0mm}{0pt}Fermionic $\eta$ (octet)
&$m^{octet}_{{\eta}_{1,2}}$&$\{0.676,\,\,1.480\}$\\\hline
\rule[4.0mm]{0mm}{0pt}Fermionic $\eta$ (singlet)
&$m^{singlet}_{{\eta}_{1,2}}$&$\{0.479,\,\,2.089\}$\\\hline
\rule[4.0mm]{0mm}{0pt}Scalar $\eta$ Higgs (octet)
&$m^{s-octet}_{\tilde{\eta}_{1,2}}$&$\{0.454,\,\,1.703\}$\\\hline
\rule[4.0mm]{0mm}{0pt}Pseudoscalar $\eta$ Higgs (octet)
&$m^{p-octet}_{\tilde{\eta}_{1,2}}$&$\{0.908,\,\,1.259\}$\\\hline
\rule[4.0mm]{0mm}{0pt}Scalar $\eta$ Higgs (singlet)
&$m^{s-singlet}_{\tilde{\eta}_{1,2}}$&$\{0.717,\,\,1.868\}$\\\hline
\rule[4.0mm]{0mm}{0pt}Pseudoscalar $\eta$  Higgs (singlet)
&$m^{p-singlet}_{\tilde{\eta}_{1,2}}$&$\{0.264,\,\,2.310\}$\\\hline
\end{tabular}
\caption{\footnotesize \small{Sparticle masses for the choice
$M_{aux}=47.112$ TeV, $\tan{\beta}=3.986$, $\mu =0.870$ TeV,
$y_{b}=0.0713$, $\lambda=0.1$,
 $\kappa=0.05$, $g_{4}=0.55$, $u=-4.024$ TeV, $M_{\eta}=1.0$ TeV and $M_t=0.176$ TeV.}}
  \label{D3}
  \end{center}
\end{table}
\end{center}
\newpage
~ \vspace{0.05in}

\begin{center}
\begin{table}[!h]
\begin{center}
\begin{tabular}{|l|c|c|}\hline
\rule[4.0mm]{0mm}{0pt} MSSM Particles&Symbol& Mass (TeV)\\\hline
\rule[4.0mm]{0mm}{0pt}Neutralinos&$\{m_{\tilde{\chi}_{1}^{0}},\,\,m_{\tilde{\chi}_{2}^{0}},\,\,m_{\tilde{\chi}_{3}^{0}},\,\,m_{\tilde{\chi}_{4}^{0}}\}$&$\{0.19625,\,\,0.585,\,\,1.179,\,\,1.184\}$\\\hline
\rule[4.0mm]{0mm}{0pt}Charginos&$\{m_{\tilde{\chi}_{1}^{\pm}},\,\,m_{\tilde{\chi}_{2}^{\pm}}\}$&$\{0.196291,\,\,1.183\}$\\\hline
\rule[4.0mm]{0mm}{0pt}Gluino&$M_{3}$&$1.411$\\\hline
\rule[4.0mm]{0mm}{0pt}Higgs boson
&$\{m_{h},\,\,m_{H},\,\,m_{A},\,\,m_{H^{\pm}}\}$&$\{0.115,\,\,1.177,\,\,1.176,\,\,1.179\}$\\\hline
\rule[4.0mm]{0mm}{0pt}R.H sleptons
&$\{m_{\tilde{e}_{R}},\,\,m_{\tilde{\mu}_{R}},\,\,m_{\tilde{\tau}_{1}}\}$&$\{0.245,\,\,0.245,\,\,0.232\}$\\\hline
\rule[4.0mm]{0mm}{0pt}L.H sleptons
&$\{m_{\tilde{e}_{L}},\,\,m_{\tilde{\mu}_{L}},\,\,m_{\tilde{\tau}_{2}}\}$&$\{0.254,\,\,0.254,\,\,0.263\}$\\\hline
\rule[4.0mm]{0mm}{0pt}Sneutrinos&$\{m_{\tilde{\nu}_{e}},\,\,m_{\tilde{\nu}_{\mu}},\,\,m_{\tilde{\nu}_{\tau}}\}$&$\{0.242,\,\,0.242,\,\,0.242\}$\\\hline
\rule[4.0mm]{0mm}{0pt}R.H down squarks
&$\{m_{\tilde{d}_{R}},\,\,m_{\tilde{s}_{R}},\,\,m_{\tilde{b}_{1}}\}$&$\{1.373,\,\,1.373,\,\,1.369\}$\\\hline
\rule[4.0mm]{0mm}{0pt}L.H down squraks
&$\{m_{\tilde{d}_{L}},\,\,m_{\tilde{s}_{L}},\,\,m_{\tilde{b}_{2}}\}$&$\{1.361,\,\,1.361\,\,1.195\}$\\\hline
\rule[4.0mm]{0mm}{0pt}R.H up squarks
&$\{m_{\tilde{u}_{R}},\,\,m_{\tilde{c}_{R}},\,\,m_{\tilde{t}_{1}}\}$&$\{1.365,\,\,1.365,\,\,0.983\}$\\\hline
\rule[4.0mm]{0mm}{0pt}L.H up squraks
&$\{m_{\tilde{u}_{L}},\,\,m_{\tilde{c}_{L}},\,\,m_{\tilde{t}_{2}}\}$&$\{1.359\,\,1.359,\,\,1.244\}$\\\hline\hline
\rule[4.0mm]{0mm}{0pt} New Particles& Symbol&Mass
(TeV)\\\hline\hline \rule[4.0mm]{0mm}{0pt}$SU(3)_{H}$ Gauge boson
octet&$M_{V}$&$1.871$\\\hline
 \rule[4.0mm]{0mm}{0pt}Singlet Higgsino
&$m_{\tilde{{\phi}}}$&$0.544$\\\hline \rule[4.0mm]{0mm}{0pt}Octet
Higgsino/gaugino
&$m_{\tilde{\phi}_{{1,2}}}$&$\{1.553,\,\,2.191\}$\\\hline
\rule[4.0mm]{0mm}{0pt}$\phi$ Higgs bosons
&$\{m_{\phi_{s}},m_{\phi_{p}},m_{\phi-octet}\}$&$\{0.247,\,\,0.840,\,\,1.955\}$\\\hline
\rule[4.0mm]{0mm}{0pt}Fermionic $\eta$ (octet)
&$m^{octet}_{{\eta}_{1,2}}$&$\{0.716,\,\,1.397\}$\\\hline
\rule[4.0mm]{0mm}{0pt}Fermionic $\eta$ (singlet)
&$m^{singlet}_{{\eta}_{1,2}}$&$\{0.529,\,\,1.890\}$\\\hline
\rule[4.0mm]{0mm}{0pt}Scalar $\eta$ Higgs (octet)
&$m^{s-octet}_{\tilde{\eta}_{1,2}}$&$\{0.421,\,\,1.699\}$\\\hline
\rule[4.0mm]{0mm}{0pt}Pseudoscalar $\eta$ Higgs (octet)
&$m^{p-octet}_{\tilde{\eta}_{1,2}}$&$\{1.031,\,\,1.098\}$\\\hline
\rule[4.0mm]{0mm}{0pt}Scalar $\eta$ Higgs (singlet)
&$m^{s-singlet}_{\tilde{\eta}_{1,2}}$&$\{0.850,\,\,1.593\}$\\\hline
\rule[4.0mm]{0mm}{0pt}Pseudoscalar $\eta$ Higgs (singlet)
&$m^{p-singlet}_{\tilde{\eta}_{1,2}}$&$\{0.247,\,\,2.189\}$\\\hline

\end{tabular}
\caption{\footnotesize \small{Sparticle masses for the choice
$M_{aux}=63.695$ TeV, $\tan{\beta}=3.999$, $\mu =1.177$ TeV,
$y_{b}=0.0716$, $\lambda=0.1$,
 $\kappa=0.08$, $g_{4}=0.55$, $u=-3.402$ TeV, $M_{\eta}=1.0$ TeV and $M_t=0.1743$ TeV.}}
\label{D4}
\end{center}
\end{table}
\end{center}

\newpage
~ \vspace{0.05in}

\begin{center}
\begin{table}[!h]
\begin{center}
\begin{tabular}{|l|c|c|}\hline
\rule[4.0mm]{0mm}{0pt} MSSM Particles&Symbol &Mass (TeV)\\\hline
\rule[4.0mm]{0mm}{0pt}Neutralinos&$\{m_{\tilde{\chi}_{1}^{0}},\,\,m_{\tilde{\chi}_{2}^{0}},\,\,m_{\tilde{\chi}_{3}^{0}},\,\,m_{\tilde{\chi}_{4}^{0}}\}$&$\{0.143,\,\,0.434,\,\,0.872,\,\,0.879\}$\\\hline
\rule[4.0mm]{0mm}{0pt}Charginos&$\{m_{\tilde{\chi}_{1}^{\pm}},\,\,m_{\tilde{\chi}_{2}^{\pm}}\}$&$\{0.143219,\,\,0.878\}$\\\hline
\rule[4.0mm]{0mm}{0pt}Gluino&$M_{3}$&$1.065$\\\hline
\rule[4.0mm]{0mm}{0pt}Higgs boson
&$\{m_{h},\,\,m_{H},\,\,m_{A},\,\,m_{H^{\pm}}\}$&$\{0.113,\,\,0.897,\,\,0.896,\,\,0.900\}$\\\hline
\rule[4.0mm]{0mm}{0pt}R.H sleptons
&$\{m_{\tilde{e}_{R}},\,\,m_{\tilde{\mu}_{R}},\,\,m_{\tilde{\tau}_{1}}\}$&$\{0.825,\,\,825,\,\,0.823\}$\\\hline
\rule[4.0mm]{0mm}{0pt}L.H sleptons
&$\{m_{\tilde{e}_{L}},\,\,m_{\tilde{\mu}_{L}},\,\,m_{\tilde{\tau}_{2}}\}$&$\{0.827,\,\,0.827,\,\,0.828\}$\\\hline
\rule[4.0mm]{0mm}{0pt}Sneutrinos&$\{m_{\tilde{\nu}_{e}},\,\,m_{\tilde{\nu}_{\mu}},\,\,m_{\tilde{\nu}_{\tau}}\}$&$\{0.823,\,\,0.823,\,\,0.823\}$\\\hline
\rule[4.0mm]{0mm}{0pt}R.H down squarks
&$\{m_{\tilde{d}_{R}},\,\,m_{\tilde{s}_{R}},\,\,m_{\tilde{b}_{1}}\}$&$\{1.017,\,\,1.017,\,\,1.014\}$\\\hline
\rule[4.0mm]{0mm}{0pt}L.H down squraks
&$\{m_{\tilde{d}_{L}},\,\,m_{\tilde{s}_{L}},\,\,m_{\tilde{b}_{2}}\}$&$\{1.008,\,\,1.008,\,\,0.886\}$\\\hline
\rule[4.0mm]{0mm}{0pt}R.H up squarks
&$\{m_{\tilde{u}_{R}},\,\,m_{\tilde{c}_{R}},\,\,m_{\tilde{t}_{1}}\}$&$\{1.011,\,\,1.011,\,\,0.718\}$\\\hline
\rule[4.0mm]{0mm}{0pt}L.H up squraks
&$\{m_{\tilde{u}_{L}},\,\,m_{\tilde{c}_{L}},\,\,m_{\tilde{t}_{2}}\}$&$\{1.005,\,\,1.005,\,\,0.944\}$\\\hline\hline
\rule[4.0mm]{0mm}{0pt} New Particles& Symbol& Mass
(TeV)\\\hline\hline \rule[4.0mm]{0mm}{0pt}$SU(3)_{H} $Gauge boson
octet&$M_{V}$&$3.779$\\\hline \rule[4.0mm]{0mm}{0pt}Singlet
Higgsino
&$m_{\tilde{{\phi}}}$&$1.058$\\
\hline \rule[4.0mm]{0mm}{0pt}Octet Higgsino/gaugino
&$m_{\tilde{\phi}_{{1,2}}}$&$\{3.071,\,\,4.495\}$\\\hline
\rule[4.0mm]{0mm}{0pt}$\phi$ Higgs bosons
&$\{m_{\phi_{s}},m_{\phi_{p}},m_{\phi-octet}\}$&$\{0.465,\,\,1.646,\,\,3.940\}$\\\hline
\rule[4.0mm]{0mm}{0pt}Fermionic $\eta$ (octet)
&$m^{octet}_{{\eta}_{1,2}}$&$\{0.254,\,\,2.521\}$\\\hline
\rule[4.0mm]{0mm}{0pt}Fermionic $\eta$ (singlet)
&$m^{singlet}_{{\eta}_{1,2}}$&$\{0.137,\,\,4.672\}$\\\hline
\rule[4.0mm]{0mm}{0pt}Scalar $\eta$ Higgs (octet)
&$m^{s-octet}_{\tilde{\eta}_{1,2}}$&$\{0.588,\,\,3.090\}$\\\hline
\rule[4.0mm]{0mm}{0pt}Pseudoscalar $\eta$ Higgs (octet)
&$m^{p-0ctet}_{\tilde{\eta}_{1,2}}$&$\{1.058,\,\,1.952\}$\\\hline
\rule[4.0mm]{0mm}{0pt}Scalar $\eta$ Higgs (singlet)
&$m^{s-singlet}_{\tilde{\eta}_{1,2}}$&$\{0.964,\,\,4.116\}$\\\hline
\rule[4.0mm]{0mm}{0pt}Pseudoscalar $\eta$ Higgs (singlet)
&$m^{p-singlet}_{\tilde{\eta}_{1,2}}$&$\{0.711,\,\,5.224\}$\\\hline

\end{tabular}
\caption{\footnotesize \small{Sparticle masses for the choice
$M_{aux}=47.112$ TeV, $\tan{\beta}=3.986$, $\mu =0.870$ TeV,
$y_{b}=0.0713$, $\lambda=0.3$,
 $\kappa=0.14$, $g_{4}=1.0$, $u=-3.779$ TeV, $M_{\eta}=0.800$ TeV and $M_t=0.176$ TeV.}}
  \label{D5}
  \end{center}
\end{table}
\end{center}

\newpage

\section{ Experimental Signatures}
  The Lightest SUSY particle in the model is the neutral Wino
  ($\chi_1^0$) which is nearly mass degenerate with the lightest
  chargino ($\chi_{1}^{\pm}$), with a mass splitting of about 235
  MeV. At the Tevatron Run 2 as well as at the LHC, the process $p\bar{p}\,\,(or\,\,pp)\rightarrow
   \chi_{1}^0+\chi_{1}^{\pm}$
  will produce these SUSY particles. Naturalness suggest that
  $m_{\chi_{1}^0}$, $m_{\chi_{1}^{\pm}}$ $\lesssim$ 300 GeV (corresponding to $m_{gluino}\lesssim 2$
  TeV). Strategies for detecting such a quasi--degenerate pair has
  been carried out in Ref. \cite{Gunion,wells,roy}. In the MSSM sector our
  model predicts $\tan{\beta}\simeq 4.0$ and $m_h\lesssim 120$ GeV, both of which can
  be tested at the LHC.

  If the $SU(3)_H$ gauge coupling $g_4$ takes small values ($g_4\simeq
  0.55$), the slepton masses will be near the current experimental
  limit. For larger values of $g_4$ ($\simeq 1.0$) the slepton
  masses are comparable to those of the squarks.

  The $SU(3)_H$ gauge boson masses are in the range $M_V=1.5-4.0$
  TeV. Although relatively light, these particles do not mediate leptonic
  FCNC, owing to the approximate $SU(3)_{H+G}$ global symmetries
  present in the model.

  The most stringent constraint on $M_V$ arises from the process
  $e^+e^-\rightarrow \mu^+\mu^-$. LEP II has set severe
  constraints on lepton compositeness \cite{Eichten,Particle} from this
  process. We focus on one such amplitude, involving all
  left--handed lepton fields. In our model, the effective
  Lagrangian for this process is
\begin{eqnarray}\label{A.77}
\textit{L}^{\rm eff}
=-\frac{2g_{4}^2}{3M_{V}^{2}}(\bar{e_{L}}\gamma_{\mu}e_{L})(\bar{\mu_{L}}\gamma^{\mu}
\mu_{L}).
\end{eqnarray}
Comparing with $\Lambda_{LL}^{-}(ee \mu \mu)>6.3$ TeV [17], we
obtain $\frac{M_V}{g_4}\geq 2.05$ TeV. For $g_4=0.55 ~(1.0)$ this
implies $M_V\geqslant$ 1.129 (2.052) TeV. From Tables 2--4 we find
that these constraints are satisfied.

The model as it stands has an unbroken $Z_2$ symmetry (in addition
to the usual $R$--parity) under which the superfields $\eta,
\bar{\eta}$ are odd and all other superfields are even.  If this
symmetry is exact, the lightest of the $\eta$ and $\bar{\eta}$
fields (a pseudoscalar singlet Higgs in the fits of Tables 2-3 and
a singlet fermion in Table 4) will be stable. We envision this
$Z_2$ symmetry to be broken by higher dimensional terms of the
type $L_\alpha H_u \Phi^\alpha \bar{\eta}_\beta
\Phi^\beta/\Lambda^2$. Such a term will induce the decay
$\eta_1^{p-singlet} \rightarrow L + \chi_1^0$ with a lifetime less
than 1 second for $\Lambda \leq 10^9$ GeV.  This would make these
$\eta$ particles cosmologically safe.  It may be pointed out that
the same effective operator, along with a TeV scale mass for the
$\eta$ fields, can provide small neutrino masses even in the
absence of the operators given in Eq. (7).

\section{Origin of the $\mu$ term}

Any satisfactory SUSY breaking model should also have a natural
explanation for the $\mu$ term (the coefficient of $H_{u} H_{d}$
term in Eq. (\ref{AN1})). In gravity mediated SUSY breaking
models, there are at least three solutions to the $\mu$ problem.
The Giudice--Masiero mechanism  \cite{Masiero} which explains the
$\mu$ term through the Kahler potential $\int
H_{u}H_{d}Z^{*}d^{4}\theta/M_{pl}$ is not readily adaptable to the
AMSB framework. The NMSSM extension which introduces singlet
fields can in principle provide a natural explanation of the $\mu$
term in the AMSB scenario. We have however found that replacing
$\mu H_{u}H_{d}$ by the term $SH_uH_d$ in the superpotential alone
can not lead to realistic SUSY breaking. It is possible to make
the NMSSM scenario compatible with symmetry breaking in the AMSB
framework by introducing a new set of fields which couple to the
singlet $S$. We do not follow this non--minimal alternative here.

There is a  natural explanation for the $\mu$ parameter in the
context of AMSB models, as suggested in  Ref. \cite{Randall}. It
assumes a Lagrangian term $L\supset
\alpha\int{d^4{\theta}\frac{(\Sigma+\Sigma^{\dag})}{M_{Pl}}H_uH_d\Phi^{\dag}\Phi}$,
where $\Sigma$ is a hidden sector field which breaks SUSY and
$\Phi$ is the compensator field. After a rescaling,
$H_u\rightarrow\Phi H_u$, $H_d\rightarrow\Phi H_d$, this term
becomes $L\supset
\alpha\int{d^4{\theta}\frac{(\Sigma+\Sigma^{\dag})}{M_{Pl}}H_uH_d\frac{\Phi^{\dag}}{\Phi}}$,
which generates a $\mu$ term in a way similar to the
Giudice--Masiero mechanism \cite{Masiero}. The $B\mu$ term is
induced only through the super--Weyl anomaly and has the form
given in Eq. (\ref{AN50}). Our predictions for $\tan\beta$ and
$m_h$ depend sensitively on this assumption.

We now point out that the $\mu$ term may have an alternative
explanation in the context of AMSB models.  This is obtained by
promoting $\mu H_{u}H_{d}$ in the superpotential to the following
\cite{mu55}:
\begin{eqnarray}\label{A.292}
W^{\prime}=\frac{aH_uH_dS^2}{M_{Pl}}+\frac{bS^2\bar{S}^2}{M_{Pl}}.
\end{eqnarray}
Here $S$ and $\bar{S}$ are standard model singlet fields.
Including AMSB induced soft parameters for these singlets (which
can arise in a variety of ways), this superpotential will have a
minimum where $\left\langle S\right\rangle\simeq \left\langle
\bar{S}\right\rangle\simeq\sqrt{M_{SUSY}M_{Pl}}$. This would
induce $\mu$ term of order $M_{SUSY}$, as needed. From the
effective low energy point of view, the superpotential will appear
to have an explicit $\mu$ term.  The $B$ term will have a form as
given in Eq. (4).

\section{ Conclusions}
In this paper we have suggested a new scenario for solving the
tachyonic slepton mass problem of anomaly mediated SUSY breaking
models. An asymptotically free $SU(3)_H$ horizontal gauge symmetry
acting on the lepton superfields provides positive masses to the
sleptons. The $SU(3)_H$ symmetry must be broken at the TeV scale.
Potentially dangerous FCNC processes mediated by the $SU(3)_H$
gauge bosons are shown to be suppressed adequately via approximate
global symmetries that are present in the model.

Our scenario predicts $m_h\lesssim$ 120 GeV for the lightest Higgs
boson mass of MSSM and $\tan{\beta}\simeq$ 4.0. The lightest SUSY
particle is the neutral Wino which is nearly degenerate with the
lightest chargino and is a candidate for cold dark matter. The
full spectrum of the model is given in Tables 2--4 for various
choices of input parameters. The very few parameters of our model
are highly constrained by the consistency of symmetry breaking.

\appendix
\section{Appendix}
 In this Appendix we give the one-loop anomalous dimension,
beta-function and the soft masses.
\subsection{Anomalous  Dimensions}
The one loop anomalous dimensions for the fields in our model are:
\begin{eqnarray}\label{A.7}
16\pi^{2}\gamma_{L_{\alpha}}&=&Y_{E_{\alpha}}^{2}-\frac{3}{10}g_{1}^{2}-\frac{3}{2}g_{2}^{2}-\frac{8}{3}g_{4}^{2},\\
16\pi^{2}\gamma_{e^{c}_{\alpha}}&=&2Y_{E_{\alpha}}^{2}-\frac{6}{5}g_{1}^{2}-\frac{8}{3}g_{4}^{2},\\
16\pi^{2}\gamma_{Q_{ij}}&=&(Y_dY_d^\dag)_{ji}+(Y_uY_u^\dag)_{ji}-\delta_i^j\left(\frac{1}{30}g_{1}^{2}+\frac{3}{2}g_{2}^{2}+\frac{8}{3}g_{3}^{2}\right),\\
16\pi^{2}\gamma_{U_{ij}}&=&2(Y_{u}^{\dag}Y_{u})_{ij}-\delta_{i}^{j}\left(\frac{8}{15}g_{1}^{2}+\frac{8}{3}g_{3}^{2}\right),\\
16\pi^{2}\gamma_{D_{ij}}&=&2(Y_{d}^{\dag}Y_{d})_{ij}-\delta_{i}^{j}\left(\frac{2}{15}g_{1}^{2}+\frac{8}{3}g_{3}^{2}\right),\\
16\pi^{2}\gamma_{H_{d}}&=&3Y_{d_{3}}^{2}-\frac{3}{10}g_{1}^{2}-\frac{3}{2}g_{2}^{2},\\
16\pi^{2}\gamma_{H_{u}}&=&3Y_{u_{3}}^{2}-\frac{3}{10}g_{1}^{2}-\frac{3}{2}g_{2}^{2},\\
16\pi^{2}\gamma_{\phi_{i}}&=&2\kappa^2+8\lambda^{2}-\frac{8}{3}g_{4}^{2},\\
16\pi^{2}\gamma_{\eta}&=&10\lambda^2-\frac{8}{3}g_{4}^{2},\\
16\pi^{2}\gamma_{\bar{\eta}}&=&-\frac{8}{3}g_{4}^{2}.
\end{eqnarray}

\subsection{Beta Function }
The beta functions for the Yukawa couplings appearing in the
superpotential, Eq. (\ref{AN1}), are:
\begin{eqnarray}\label{A.13}
\beta(Y_{d_{3}})&=&\frac{Y_{d_{3}}}{16\pi^2}\left(6Y_{d_{3}}^2+Y_{u_{3}}^2
-\frac{7}{15}g_{1}^{2}-3g_{2}^{2}-\frac{16}{3}g_{3}^{2}\right),\\
\beta(Y_{u_{3}})&=&\frac{Y_{u_{3}}}{16\pi^2}\left(6Y_{u_{3}}^2+Y_{d_{3}}^2
-\frac{13}{15}g_{1}^{2}-3g_{2}^{2}-\frac{16}{3}g_{3}^{2}\right),\\
\beta(Y_{E_{\alpha}})&=&\frac{Y_{E_{\alpha}}}{16\pi^2}\left(4Y_{E_{\alpha}}^2+3Y_{d_{3}}^2
-\frac{9}{5}g_{1}^{2}-3g_{2}^{2}\right),\\
\beta\left(\lambda\right)&=&\frac{\lambda}{16\pi^2}\left(28\lambda^{2}+2\kappa^{2}
-8g_{4}^{2}\right),\\
\beta\left(\kappa\right)&=&\frac{3\kappa}{16\pi^2}\left(2\kappa^{2}+8\lambda^{2}
-\frac{8}{3}g_{4}^{2}\right).
\end{eqnarray}

The gauge beta function of our model are
\begin{eqnarray}\label{A.19}
\beta(g_{i})&=&b_{i}\frac{g_{i}^{3}}{16\pi^2},
\end{eqnarray}
where $b_{i}=(\frac{33}{5}, 1, -3, -3)$ for $i=1-4$.
\subsection{$A$ terms}
The trilinear soft SUSY breaking terms are given by
\begin{eqnarray}\label{A.222}
A_{Y}&=&-\frac{\beta{(Y)}}{Y}M_{aux},
\end{eqnarray}
where $Y=(Y_{u_i}, Y_{d_i},Y_{E_{\alpha}} ,k,\lambda$).
\subsection{Gaugino  Masses}
The soft masses of the gauginos are given by:
\begin{eqnarray}\label{A.223}
M_{i}&=&\frac{\beta{(g_{i})}}{g_{i}}M_{aux},
\end{eqnarray}
where $i=1,2,3,4$,  corresponding to the gauge groups $U(1)_{Y}$,
$SU(2)_{W}$, $SU(3)_{C}$ and $SU(3)_{H}$, with $\beta(g_i)$ given
as in Eq. (55).

\subsection{Soft SUSY Masses} The soft masses of the squarks
and the sleptons are given in the text. For the $H_u$, $H_d$
$\Phi_i$, $\eta_i$, $\bar{\eta}$ fields they are:
\begin{eqnarray}\label{A.291}
(\tilde{m}^{2}_{soft})_{H_{u}}^{H_{u}}&=&\frac{M_{aux}^{2}}{16\pi^2}\left(3Y_{u_{3}}
\beta{(Y_{u_{3}})}-\frac{3}{10}g_{1}\beta{(g_{1})}-\frac{3}{2}g_{2}\beta{(g_{2})}\right),\\
(\tilde{m}^{2}_{soft})_{H_{d}}^{H_{d}}&=&\frac{M_{aux}^{2}}{16\pi^2}\left(3Y_{d_{3}}
\beta{(Y_{d_{3}})}-\frac{3}{10}g_{1}\beta{(g_{1})}-\frac{3}{2}g_{2}\beta{(g_{2})}\right),\\
(\tilde{m}^{2}_{soft})_{\Phi_{i}}^{\Phi_{i}}&=&\frac{M_{aux}^{2}}{16\pi^2}\left(2\kappa
\beta{(\kappa)}+8\lambda\beta{(\lambda)}-\frac{8}{3}g_{4}\beta{(g_{4})}\right),\\
(\tilde{m}^{2}_{soft})_{\eta}^{\eta}&=&\frac{M_{aux}^{2}}{16\pi^2}\left(10\lambda\beta{(\lambda)}-\frac{8}{3}g_{4}\beta{(g_{4})}\right),\\
(\tilde{m}^{2}_{soft})_{\bar{\eta}}^{\bar{\eta}}&=&\frac{M_{aux}^{2}}{16\pi^2}\left(-\frac{8}{3}g_{4}\beta{(g_{4})}\right).
\end{eqnarray}

\section*{Acknowledgments}
This work is supported in part by DOE Grant \# DE-FG03-98ER-41076,
an award from the Research Corporation and by DOE Grant \#
DE-FG02-01ER-45684. The work of I.G. was supported in part by the
National Science Foundation under grant PHY00-98791.

\end{document}